\documentstyle[12pt,amssymb]{article}

\begin{document}

\newtheorem{definition}{Definition $\!\!$}[section]
\newtheorem{prop}[definition]{Proposition $\!\!$}
\newtheorem{lem}[definition]{Lemma $\!\!$}
\newtheorem{corollary}[definition]{Corollary $\!\!$}
\newtheorem{theorem}[definition]{Theorem $\!\!$}
\newtheorem{example}[definition]{\it Example $\!\!$}
\newtheorem{remark}[definition]{Remark $\!\!$}

\newcommand{\nc}[2]{\newcommand{#1}{#2}}
\newcommand{\rnc}[2]{\renewcommand{#1}{#2}}
\nc{\bpr}{\begin{prop}}
\nc{\bth}{\begin{theorem}}
\nc{\ble}{\begin{lem}}
\nc{\bco}{\begin{corollary}}
\nc{\bre}{\begin{remark}}
\nc{\bex}{\begin{example}}
\nc{\bde}{\begin{definition}}
\nc{\ede}{\end{definition}}
\nc{\epr}{\end{prop}}
\nc{\ethe}{\end{theorem}}
\nc{\ele}{\end{lem}}
\nc{\eco}{\end{corollary}}
\nc{\ere}{\hfill\mbox{$\Diamond$}\end{remark}}
\nc{\eex}{\end{example}}
\nc{\epf}{\hfill\mbox{$\Box$}}
\nc{\ot}{\otimes}
\nc{\bsb}{\begin{Sb}}
\nc{\esb}{\end{Sb}}
\nc{\ct}{\mbox{${\cal T}$}}
\nc{\ctb}{\mbox{${\cal T}\sb B$}}
\nc{\bcd}{\[\begin{CD}}
\nc{\ecd}{\end{CD}\]}
\nc{\ba}{\begin{array}}
\nc{\ea}{\end{array}}
\nc{\bea}{\begin{eqnarray}}
\nc{\eea}{\end{eqnarray}}
\nc{\be}{\begin{enumerate}}
\nc{\ee}{\end{enumerate}}
\nc{\beq}{\begin{equation}}
\nc{\eeq}{\end{equation}}
\nc{\bi}{\begin{itemize}}
\nc{\ei}{\end{itemize}}
\nc{\kr}{\mbox{Ker}}
\nc{\te}{\!\ot\!}
\nc{\pf}{\mbox{$P\!\sb F$}}
\nc{\pn}{\mbox{$P\!\sb\nu$}}
\nc{\bmlp}{\mbox{\boldmath$\left(\right.$}}
\nc{\bmrp}{\mbox{\boldmath$\left.\right)$}}
\rnc{\phi}{\mbox{$\varphi$}}
\nc{\LAblp}{\mbox{\LARGE\boldmath$($}}
\nc{\LAbrp}{\mbox{\LARGE\boldmath$)$}}
\nc{\Lblp}{\mbox{\Large\boldmath$($}}
\nc{\Lbrp}{\mbox{\Large\boldmath$)$}}
\nc{\lblp}{\mbox{\large\boldmath$($}}
\nc{\lbrp}{\mbox{\large\boldmath$)$}}
\nc{\blp}{\mbox{\boldmath$($}}
\nc{\brp}{\mbox{\boldmath$)$}}
\nc{\LAlp}{\mbox{\LARGE $($}}
\nc{\LArp}{\mbox{\LARGE $)$}}
\nc{\Llp}{\mbox{\Large $($}}
\nc{\Lrp}{\mbox{\Large $)$}}
\nc{\llp}{\mbox{\large $($}}
\nc{\lrp}{\mbox{\large $)$}}
\nc{\lbc}{\mbox{\Large\boldmath$,$}}
\nc{\lc}{\mbox{\Large$,$}}
\nc{\Lall}{\mbox{\Large$\forall$}}
\nc{\bc}{\mbox{\boldmath$,$}}
\rnc{\epsilon}{\varepsilon}
\rnc{\ker}{\mbox{\em Ker}}
\nc{\ra}{\rightarrow}
\nc{\ci}{\circ}
\nc{\cc}{\!\ci\!}
\nc{\T}{\mbox{\sf T}}
\nc{\can}{\mbox{\em\sf T}\!\sb R}
\nc{\cnl}{$\mbox{\sf T}\!\sb R$}
\nc{\lra}{\longrightarrow}
\nc{\M}{\mbox{Map}}
\rnc{\to}{\mapsto}
\nc{\imp}{\Rightarrow}
\rnc{\iff}{\Leftrightarrow}
\nc{\bmq}{\cite{bmq}}
\nc{\ob}{\mbox{$\Omega\sp{1}\! (\! B)$}}
\nc{\op}{\mbox{$\Omega\sp{1}\! (\! P)$}}
\nc{\oa}{\mbox{$\Omega\sp{1}\! (\! A)$}}
\nc{\inc}{\mbox{$\,\subseteq\;$}}
\nc{\de}{\mbox{$\Delta$}}
\nc{\spp}{\mbox{${\cal S}{\cal P}(P)$}}
\nc{\dr}{\mbox{$\Delta_{R}$}}
\nc{\dsr}{\mbox{$\Delta_{\cal R}$}}
\nc{\m}{\mbox{m}}
\nc{\0}{\sb{(0)}}
\nc{\1}{\sb{(1)}}
\nc{\2}{\sb{(2)}}
\nc{\3}{\sb{(3)}}
\nc{\4}{\sb{(4)}}
\nc{\5}{\sb{(5)}}
\nc{\6}{\sb{(6)}}
\nc{\7}{\sb{(7)}}
\nc{\hsp}{\hspace*}
\nc{\nin}{\mbox{$n\in\{ 0\}\!\cup\!{\Bbb N}$}}
\nc{\al}{\mbox{$\alpha$}}
\nc{\bet}{\mbox{$\beta$}}
\nc{\ha}{\mbox{$\alpha$}}
\nc{\hb}{\mbox{$\beta$}}
\nc{\hg}{\mbox{$\gamma$}}
\nc{\hd}{\mbox{$\delta$}}
\nc{\he}{\mbox{$\varepsilon$}}
\nc{\hz}{\mbox{$\zeta$}}
\nc{\hs}{\mbox{$\sigma$}}
\nc{\hk}{\mbox{$\kappa$}}
\nc{\hm}{\mbox{$\mu$}}
\nc{\hn}{\mbox{$\nu$}}
\nc{\la}{\mbox{$\lambda$}}
\nc{\hG}{\mbox{$\Gamma$}}
\nc{\hD}{\mbox{$\Delta$}}
\nc{\th}{\mbox{$\theta$}}
\nc{\Th}{\mbox{$\Theta$}}
\nc{\ho}{\mbox{$\omega$}}
\nc{\hO}{\mbox{$\Omega$}}
\nc{\hp}{\mbox{$\pi$}}
\nc{\hP}{\mbox{$\Pi$}}
\nc{\bpf}{{\it Proof.~~}}
\nc{\slq}{\mbox{$A(SL\sb q(2))$}}
\nc{\fr}{\mbox{$Fr\llp A(SL(2,\IC))\lrp$}}
\nc{\slc}{\mbox{$A(SL(2,\IC))$}}
\nc{\af}{\mbox{$A(F)$}}

\def\esl{{\mbox{$E\sb{\frak s\frak l (2,{\Bbb C})}$}}}
\def\esu{{\mbox{$E\sb{\frak s\frak u(2)}$}}}
\def\zf{{\mbox{${\Bbb Z}\sb 4$}}}
\def\zt{{\mbox{$2{\Bbb Z}\sb 2$}}}
\def\ox{{\mbox{$\Omega\sp 1\sb{\frak M}X$}}}
\def\oxh{{\mbox{$\Omega\sp 1\sb{\frak M-hor}X$}}}
\def\oxs{{\mbox{$\Omega\sp 1\sb{\frak M-shor}X$}}}

\renewcommand{\thesection}{\arabic{section}}

\nc{\Section}{\setcounter{definition}{0}\section}

\newcounter{c}
\renewcommand{\[}{\setcounter{c}{1}$$}
\newcommand{\etyk}[1]{\vspace{-7.4mm}$$\begin{equation}\label{#1}
\addtocounter{c}{1}}
\renewcommand{\]}{\ifnum \value{c}=1 $$\else \end{equation}\fi}

\newcommand{\dowod}{\noindent{\bf Proof:} }
\newcommand{\Sp}{{\rm Sp}\,}
\newcommand{\Mor}{\mbox{$\rm Mor$}}
\newcommand{\skrA}{{\cal A}}
\newcommand{\Phase}{\mbox{$\rm Phase\,$}}
\newcommand{\id}{{\rm id}}
\newcommand{\diag}{{\rm diag}}
\newcommand{\inv}{{\rm inv}}
\newcommand{\ad}{{\rm ad}}
\newcommand{\poi}{{\rm pt}}
\newcommand{\Dim}{{\rm dim}\,}
\newcommand{\Ker}{{\rm ker}\,}
\newcommand{\Mat}{{\rm Mat}\,}
\newcommand{\Rep}{{\rm Rep}\,}
\newcommand{\Fun}{{\rm Fun}\,}
\newcommand{\Tr}{{\rm Tr}\,}
\newcommand{\supp}{\mbox{$\rm supp$}}
\newcommand{\half}{\frac{1}{2}}

\newcommand{\skrF}{{A}}
\newcommand{\skrD}{{\cal D}}
\newcommand{\skrC}{{\cal C}}

\newcommand{\ttimes}{\mbox{$\hspace{.5mm}\bigcirc\hspace{-4.9mm}
\perp\hspace{1mm}$}}
\newcommand{\Ttimes}{\mbox{$\hspace{.5mm}\bigcirc\hspace{-3.7mm}
\raisebox{-.7mm}{$\top$}\hspace{1mm}$}}
\newcommand{\Cstar}{$^*$-}

\newcommand{\bbr}{{\bf R}}
\newcommand{\bbz}{{\bf Z}}
\newcommand{\Ci}{C_{\infty}}
\newcommand{\Cb}{C_{b}}
\newcommand{\fa}{\forall}
\newcommand{\rrr}{right regular representation}
\newcommand{\wrt}{with respect to}

\newcommand{\qg}{quantum group}
\newcommand{\qgs}{quantum groups}
\newcommand{\cs}{classical space}
\newcommand{\qs}{quantum space}
\newcommand{\po}{Pontryagin}
\newcommand{\ch}{character}
\newcommand{\chs}{characters}

\def\inbar{\,\vrule height1.5ex width.4pt depth0pt}
\def\IC{{\Bbb C}}
\def\IZ{{\Bbb Z}}
\def\IN{{\Bbb N}}
\def\IH{{\Bbb H}}
\def\otc{\otimes_{\IC}}
\def\ra{\rightarrow}
\def\ota{\otimes_ A}
\def\otza{\otimes_{ Z(A)}}
\def\otc{\otimes_{\IC}}
\def\h{\rho}
\def\x{\zeta}
\def\th{\theta}
\def\s{\sigma}
\def\t{\tau}
\def\st{\stackrel}
\def\Fr{\mbox{Fr}}
\def\gal{-Galois extension}
\def\coq{$\cal H$}
\title{\vspace*{-0.0mm}{\Large\bf A FINITE QUANTUM SYMMETRY OF 
$M(3,\IC )$}}
\author{Ludwik D\c{a}browski$\!\!$
\thanks{
e-mail: dabrow@sissa.it, ~$^{\dag}$e-mail: nesti@sissa.it, ~$^{\ddag}$e-mail: sinis@sissa.it}~,
~Fabrizio Nesti$^{\dag}$,
~Pasquale Siniscalco$^{\ddag}$
\\
\normalsize SISSA, Via Beirut 2-4, Trieste, \mbox{34014}~Italy.
}
\date{}
\maketitle
\thispagestyle{empty}
\setcounter{page}{0}
\vspace{1cm}

\begin{abstract}
\hspace{.3in}\begin{minipage}{4.7in}
The 27-dimensional Hopf algebra \af, 
defined by the exact sequence of quantum groups 
$\slc\st{Fr}{\lra}\slq\st{\pi\sb F}{\lra}\af$, $q=e^{\frac{2\pi i}{3}}$,
is studied as a finite quantum group symmetry of the matrix algebra $M(3,\IC )$, 
describing the color sector of Alain Connes' formulation of the Standard Model.
The duality with the Hopf algebra $\cal H$, 
investigated in a recent work by Robert Coquereaux, is established and used 
to define a representation of $\cal H$ on $M(3,\IC )$
and two commuting representations of $\cal H$ on \af.
\end{minipage}
\end{abstract}

\vskip 7cm
\centerline{SISSA 63/97/FM}

\newpage
\Section*{Introduction}

In recent years there has been a growing interest 
in establishing the links between  
noncommutative geometry and quantum groups
in analogy to the important relations between 
classical geometry and group theory.
An interesting territory to test their possible interplay is 
Connes' formulation of the Standard Model of elementary particles 
(for its latest version, including also the gravity, see \cite{Ch-C}).
In the Standard Model, which is extraordinarily successful,
there remain still some fundamental open questions.
It is tempting to investigate if for resolving them 
some new symmetry of the quantum group type
(perhaps finite) could be helpful.
This seems quite a natural question having at a disposal 
a noncommutative formulation of the Standard Model.

At the end of Ref. \cite{c-a} the sequence 
$1\ra F\ra SU_q(2)\ra SU(2)\ra 1 $, where 
$q=e^{\frac{2\pi i}{3}}$ and $F$ is a finite quantum group,
has been suggested in relation to the Standard Model 
or its hypothetical extensions to higher energy regimes.
In the paper \cite{dhs} a possible interpretation of this sequence is given 
both in the sense of exact sequence of Hopf algebras and in the 
language of principal quantum bundles (Hopf-Galois extensions). 
More precisely, for $q^3=1$, there is an exact sequence of Hopf algebras 
$\slc\st{Fr}{\lra}\slq\st{\pi\sb F}{\lra}\af$,
where \af~ is a finite dimensional quotient Hopf algebra of \slq. 
It is furthermore shown that \slq~ is a faithfully flat \af-extension of
\slc.

In the present communication, we elaborate more on \af, 
focusing our attention on its possible role as a quantum symmetry of the 
algebra $M(3,\IC )$, which is a part of the algebra 
${\cal A} = M(3,\IC )\oplus \IH \oplus \IC$
 used by Connes for his formulation of the Standard Model. 
In section 1, we recall the basic results from \cite{dhs}. 
The first part of this section contains  
noncommutative generalizations of some geometrical structures,
and serves to give a more mathematical framework
to the main object of our interest, \af, 
which is introduced in the second part.
In section 2, we discuss the coaction of \af~ on $M(3,\IC)$, 
the color sector of ${\cal A}$, and possible
extensions to the other two sectors.
In section 3, we describe the parallel work of \cite{c-r} 
in the framework of universal enveloping algebras 
and we establish the Hopf duality between the
algebra \coq, defined therein, and \af. 
In section 4, using this duality we exhibit a representation of \coq~ on
$M(3,\IC)$ that can be extended to ${\cal A}$.
In this representation the generator $K$ of \coq~ acts as automorphism
while the generators $X_{\pm}$ act as {\it twisted} derivations.
The automormorphism is clearly inner and it turns out that the twisted 
derivations can be also expressed as a sort of internal operations 
($\IZ_3$-graded bracket with some elements $\widetilde{ X}_\pm$). 
In section 5, we give two commuting representations of \coq~ on \af.
(The explicit results are tabulated in the Appendix).
Furthermore, we use the notion of integrals 
{\it in} and {\it on} a Hopf algebra to say more about the algebraic and 
coalgebraic properties of \coq~ and \af.  
   
We consider the present contribution as a step in the direction
of achieving physically significant statements 
about a possible quantum symmetry behind the Standard Model.
In this respect it is particularly interesting
to answer the question of how to implement the finite quantum symmetry $A(F)$
on the level of representation spaces of ${\cal A}$
(which describe matter fields),
and to verify if the generators of \coq~ implemented as operators 
preserve the action integral in some sense. These questions shall be
investigated in our future work.

\section{Preliminaries}\parindent0pt

Let us recall that, given a sequence of groups and group homomorphisms  
$G\ra G' \ra G"$, one can consider the (dual) sequence $B\st{j}{\ra} P\st{\pi}{\ra} H$
of Hopf algebras of functions and of Hopf algebra morphisms
(pull back of the mappings reverses the arrows).
The exactness of the sequence of groups translates then into the definition 
\cite{pw, s1} of {\em exact} sequence of Hopf algebras.
This definition requires that $j$ has to be injective, $H=P\slash Pj(B\sp+)P$, 
where $B\sp+$ denotes  
the kernel of the counit map of $B$, and that $\pi$ is the canonical surjection.
It applies directly to the noncommutative case.

In the classical (commutative) case an exact sequence of groups 
is equivalent to a principal fibre bundle 
with principal space $G'$, base space $G'/G=G"$ and structure group $G$.
In the noncommutative case this is no longer automatically 
true, and further conditions must be imposed in order 
to achieve the equivalence.
Let us comment on this in more detail.

An exact sequence of Hopf algebras is 
called {\em strictly exact} \cite{s1} iff 
$P$ is a right faithfully flat module over $j(B)$,
and $j(B)$ is a normal Hopf subalgebra of $P$, i.e., 
$p\1 j(B)S(p\2)\cup S(p\1) j(B)p\2\inc j(B)$ for any $p\in P$,
where the Sweedler notation is used for the coproduct and $S$ is the antipode.
 
Next, a suitable dualization of (some of) 
the properties of a principal fibre bundle
is achieved via the notion of Hopf-Galois extension \cite{kt}.
If $H$ is a Hopf algebra, $P$ is a right $H$-comodule algebra and 
$B=P\sp{co H}$ ~(the space of coinvariants of the coaction), 
we say that P is a (right) {\em Hopf-Galois} $H$-extension iff 
the canonical left $P$-module right $H$-comodule map 
\[
(m\sb P\ot id)\circ (id\ot\sb B \dr )\, :\; P\ot\sb B P\lra P\ot H
\]
is bijective. 
In addition, we say that it is {\em faithfully flat} iff 
$P$ is faithfully flat (right) $B$-module.

In the classical case, $B$, the algebra of functions 
on the quotient space, is identified with the subalgebra of functions on the
principal space that are constant on the fibres and the canonical map is just the
pull-back of the map $X\times G \ra X\times_M X$ given by $(x,g)\mapsto (x,xg)$,
whose bijectivity means that the action is free an transitive on the fibres. 
A particular kind of Hopf-Galois extensions is given by the 
{\it cleft} ones: 
an $H$\gal\ is called {\em cleft} iff there exists a unital, convolution
invertible, linear map $\Phi : H\ra P$ satisfying 
$\dr\circ\Phi=(\Phi\ot id)\circ\hD$.
 
In the datum of an exact sequence of Hopf algebras we have, in particular, 
a quotient Hopf algebra $H$, which coacts in a natural way via {\it push out} 
$(\Delta_R=(Id\ot \pi)\circ\Delta)$ on $P$, and a Hopf subalgebra $j(B)$. 
It remains to be verified whether $j(B)$ coincides with the space of 
coinvariants $P^{co H}$ and whether $P$ is faithfully flat Hopf-Galois $H$-extension 
of $j(B)$.
It turns out \cite{s1} that for {\em strictly exact} sequences of Hopf
algebras 
this is indeed the case.

Now we pass to the case of our interest.
We recall that $A(SL_q(2))$ is a complex Hopf algebra
generated by $T_{ij} = {\scriptsize \left( \ba{cc} a & b \\ c & d \ea
\right)}$, satisfying the 
following relations: 
\bea \label{comm}
&& ab=q ba~,~~ ac=qca~,~~ bd=qdb~,~~ bc = cb~,~~  cd =qdc~, \nonumber \\ 
&& ad-da=(q-q^{-1})bc~, ~~ ad-qbc=da-q^{-1 }bc=1\, .  
\eea
The comultiplication $\Delta$, counit $\varepsilon$, and antipode $S$ are 
\bea
&
\Delta {\scriptsize \left( \ba{cc} a & b \\ c & d \ea \right)} =
{\scriptsize \left( \ba{cc} a & b \\ c & d \ea \right)} \ot 
{\scriptsize \left( \ba{cc} a & b \\ c & d \ea \right)}~,  \\
&
\varepsilon {\scriptsize \left( \ba{cc} a & b \\ c & d \ea \right)}=
{\scriptsize \left( \ba{cc} 1 & 0 \\ 0 & 1 \ea \right)}\, ,\;\;
S{\scriptsize \left( \ba{cc} a & b \\ c & d \ea \right)}=
{\scriptsize \left( \ba{cc} d &-q^{-1} b \\ -qc & a \ea \right)}\, .\nonumber
\eea
As a complex vector space, $A(SL_q(2))$ has a basis 
$a^ib^jc^k$; $i,j,k\geq 0$ and $b^ic^jd^k$; $i,j\geq 0,k>0$.\\
 From now on, unless stated differently, 
we set the parameter $q=e^{\frac{2\pi i}{3}}$.\\ 
Let $\bar{T}\sb{ij}$ denote the generators of undeformed $A(SL(2,\IC))$.\\
In \cite{dhs} the following commutative diagram of algebras and algebra homomorphisms
has been introduced:
\beq
\label{diag}
\begin{array}{ll}
 {A}(\IC^2) \!\!\! &\!\!\!\! \mathop{-\hspace{-6pt}
\longrightarrow}\limits^{\rho} {A}(\IC^2) \otimes {A}(SL(2,\IC))  \\
\ ^{fr} \Big\downarrow 
 &  ~~~~~~~~~~~~~~~\Big\downarrow \phantom{\cdot} ^{fr\otimes F\!r} \\ 
\ {A}(\IC_q^2) \!\!\! &\!\!\!\! \mathop{-\hspace{-6pt}
\longrightarrow}\limits^{\rho_q} 
{A}(\IC^2_q) \otimes {A}(SL_q(2,\IC)) \\
\ ^{\pi\sb M} \Big\downarrow ~
 &  ~~~~~~~~~~~~~~~\Big\downarrow \phantom{\cdot} ^{\pi\sb M\otimes 
\pi\sb F} \\ \ M(3,\IC) \! &\!\!\!\! \mathop{-\hspace{-6pt}
\longrightarrow}\limits^{\rho_F} M(3,\IC) \otimes {A}(F)  \ .
\end{array}
\eeq
We explain now the various ingredients of (\ref{diag}).\\
First, ${A}(\IC^2)=\IC[\bar{x},\bar{y}]$ is the algebra of polynomials 
on $\IC\sp2$, 
${A}(\IC_q^2)$ is the algebra of the quantum plane,
i.e. the free algebra generated by $x$ and $y$ modulo the relation $xy = q yx$, 
and $fr$ is the injection given by $fr(\bar{x})=x\sp3,\; fr(\bar{y})=y\sp3$.
The map $\pi\sb M$ is the composition of the canonical projection
$x \mapsto\tilde x$, $y \mapsto \tilde y$,
from the algebra ${A}(\IC_q^2)$ to the algebra 
${A}(\IC_q^2)$ modulo the relations $x\sp3 = 1$ and $y\sp3 = 1$,
with the map
\beq
\label{weyl}
\tilde x \mapsto \pmatrix{0 & 1\sb{n-1}\cr 1 & 0} ~ , \quad
\tilde y \mapsto diag(1, q,..., q\sp{n-1})~,
\eeq
identifying (for $n=3$)
the latter algebra with the algebra of matrices $M(n,\IC)$
(see Section~IV.D.15 of~\cite{w-h}).\\
Next, the Hopf algebra injection 
$
Fr:\slc\ni \bar{T}\sb{ij}\longmapsto (T\sb{ij})\sp3\in\slq\, ,\;\; i,j = 1, 2
$,
is the so-called {\it Frobenius} mapping.
Moreover, \af~ is the finite dimensional quotient Hopf algebra of \slq ~modulo the 
relations
\beq\label{F} 
a^3 = 1 = d^3, ~~ b^3 = 0 = c^3, 
\eeq
and $\pi_F$ is the canonical projection.\\
Finally, $\rho $, $\rho_q $ and $\rho_F $ are the natural right coactions 
on  ${A}(\IC^2)$,  ${A}(\IC_q^2)$, and $M(3,\IC)$, respectively,
given symbolically by 
$e\sb i\to\sum\sb{j=1, 2} e\sb j\ot M\sb{ji}\, ,\; i=1, 2$.

In \cite{dhs} the following facts are shown:\\
i) the sequence 
$\slc\st{Fr}{\lra}\slq\st{\pi\sb F}{\lra}\af$
is strictly exact\\
ii)\slq ~is a (right, faithfully flat) Hopf-Galois \af-extension of \fr \\ 
iii) \af ~is a 27-dimensional complex vector space with a basis 
 $\tilde{a}^p \tilde{b}^r\tilde{c}^s$; $p,r,s\in\{0,1,2\}$, 
where $\tilde{a}=\pi_F(a)$, {\it etc.}\\
iv) \af ~has a faithful representation
\footnote{For a representation in terms of Grassman variables see \cite{sa}.}
 $\varrho$
\beq \varrho(\tilde{a}) =  {\bf J} \ot {\bf 1}\sb3 \ot {\bf 1}\sb3~, \quad
      \varrho(\tilde{b})  =  {\bf Q} \ot {\bf N} \ot {\bf 1}\sb3~, \quad
\varrho(\tilde{c})  = {\bf Q} \ot {\bf 1}\sb3 \ot {\bf N}~, \eeq
 where 
\beq \bf{J} =  \left( \ba{ccc} 0 & 0 & 1 \\ 1 & 0 & 0 \\ 0  & 1 & 0 \ea 
              \right)~,~~
     \bf{Q} = \left( \ba{ccc} 1 & 0 & 0 \\ 0 & q^2 & 0 \\ 0  & 0 & q \ea 
   \right)~,~~
\bf{N} =\left( \ba{ccc} 0 & 0 & 0 \\ 1 & 0 & 0 \\ 0  & 1 & 0 \ea 
\right)~.
\label{rep}
\eeq 
v) $fr(A(\IC\sp2))=A(\IC\sp2\sb q)\sp{co A(F)}$, ie.
``Frobenius-like" map $fr$ allows to identify
$A(\IC\sp2)$ with the subalgebra of 
$(id\ot\pi\sb F)\circ\rho\sb q$-coinvariants of $A(\IC\sp2\sb q)$.

\section{Quantum symmetries of $M(3,\IC )$}
First we elaborate more on $F$ as a quantum-group symmetry of $M(3,\IC)$
--- a direct summand of Connes' algebra for the Standard Model.

It is easy to see that the subalgebra of coinvariants of $M(3,\IC)$ under
the coaction of \af~ is one dimensional: $M(3,\IC)^{co A(F)}=\IC$. This 
leads us to think of $M(3,\IC)$ as a quantum homogeneous space of \af. 
Notice, however, that $M(3,\IC)$ is {\em not} an embeddable \af-space 
in the sense of \cite{bre}, because there does not exist an algebra
injection 
$i: M(3,\IC) \ra \af$ (in particular, the algebra $M(3,\IC)$ has no
characters).

The classical subgroup of $F$ is given by the set of characters of \af,
i.e. non zero algebra morphisms
$\chi:\af\ra\IC $, endowed with the convolution product
$(\chi\cdot\psi)(u)=(\chi\otimes\psi)\circ\Delta(u)$. 
It is easy to see that there are only three characters $\chi_i$, $i = 0, 1, 2$.
Their action on generators of \af ~is given by 
$\chi_i {\scriptsize \left( \ba{cc} a & b \\ c & d \ea \right)}=
{\scriptsize \left( \ba{cc} q^i & 0 \\ 0 & q^{2i} \ea \right)}$. 
The classical subgroup of $F$ is then isomorphic to
$\IZ_3$. The Hopf algebra $A(\IZ_3)$ appears as a quotient 
of \af~ by the ideal generated by $\tilde{b}$, $\tilde{c}$. 
Notice that this ideal is the intersection of the kernels of the characters.

Next, $A(\IZ_3)$ coacts on $M(3,\IC)$ via push-out. Using the basis of 
$M(3,\IC)$ given by $\tilde x^r\tilde y ^s~;~r,s\in\{0,1,2\}$, one has 
$M(3,\IC)^{co A(\IZ_3)}= 
{\mbox {\rm span}}_{\IC}\{1, \tilde x\tilde y, \tilde x^2\tilde y^2\} \cong \IC^3$. 
It turns out that the extension $(M(3,\IC), \IC^3, A(\IZ_3))$ is a cleft 
Hopf-Galois extension, with the unital, right covariant, convolution
invertible map $\Phi :A(\IZ_3)\ra M(3,\IC)$ being given by
$\Phi(\tilde{a})=\tilde
x$, $\Phi (\tilde{a}^2 )=\tilde x^2$. Being $\IC^3$ a commutative
subalgebra,
it also follows from \cite{s2} that the extension is faithfully 
flat. \footnote{
This result holds in general for any root of unity $q^n=1$, 
so that $(M(n,\IC), \IC^n, A(\IZ_n))$ 
is a cleft, faithfully flat, Hopf-Galois extension.}

Making use of the coaction of $A(\IZ_3)$ on $M(3,\IC)$ and of the
characters of $A(\IZ_3)$, we can define three algebra endomorphisms 
of $M(3,\IC)$ by 
\beq 
F_i =(Id\ot\chi_i)\circ\Delta_R~,~~ i=0,1,2~. 
\eeq   
Explicitly one has:
\beq
F_i(\tilde x) = q^i \tilde x~,~~~F_i(\tilde y) = q^{2i} \tilde y~.
\eeq
The mapping $\chi_i\mapsto F_i$ identifies 
$\IZ_3$ as a subgroup of the group of algebra automorphisms of $M(3,\IC)$,
which is $SU(3)/\IZ_{3}^{\rm diag}$, where 
$\IZ_{3}^{\rm diag} = \{ {\bf 1}_3, q{\bf 1}_3, q^2{\bf 1}_3 \}$.
More precisely, eg. the generator $\chi_1$ of $\IZ_{3}$ 
corresponds to an inner automorphism via the adjoint action 
(of the $\IZ_{3}^{\rm diag}$-class) of the matrix
\beq \label{matrice}
{\bf U}_1 = \tilde{x}^2\tilde{y}^2= \left( \ba{ccc} 0 & 0 & q \\ 1 & 0 &
0\\ 0 & q^2 & 0 \ea
\right)~.
\eeq
Hence the quantum finite symmetry $A(F)$ has $\IZ_3$ as an
overlap  with the classical symmetry group 
$SU(3)/\IZ_{3}^{\rm diag}$ of $M(3,\IC)$.

One may wonder if any nontrivial finite symmetry of the remaining piece 
$\IH \oplus \IC$ of Connes' algebra for the Standard Model
can be also obtained in the same spirit. 
As far as the algebra of quaternions $\IH$ is concerned, it embeds, at
least as a real algebra, into
$M(2,\IC)$ via the mapping
\beq
\label{IH}
u=\alpha +\beta j \mapsto 
\left(\begin{array}{cc} \alpha & \beta \\ -\bar \beta & \bar \alpha
\end{array} \right) ~, ~~\alpha,\beta \in \IC~.
\eeq
In terms of the basis of $M(2,\IC)$ given by $\tilde x^r\tilde y
^s~;~r,s\in\{0,1\}$, a quaternion $u$ is then expressed as 
$u=\frac{1}{2}\left( (\alpha +\bar\alpha)1 +(\beta-\bar\beta)\tilde x
+ (\alpha - \bar\alpha)\tilde y - (\beta+\bar\beta)\tilde x\tilde y
\right)$.\\
Then, using (\ref{weyl}) for $n=2$, we can define a coaction $ \rho_2$ on
$M(2,\IC)$ of the
quotient Hopf algebra of \slq, $q= -1$, modulo the relations 
\cite{tm}
\beq\label{F2} 
a^2 = 1 = d^2, ~~ b = 0 = c ~,
\eeq
obtaining, however, nothing but a (classical) $A(\IZ_2)$.\\
Quaternions are a (real) subcomodule of $M(2,\IC)$, since one has 
\[
\rho_2(u) =  
\left( Re (\alpha) + Re(\beta) j\right) \ot 1+ 
\left( Im (\alpha) + Im  (\beta) j\right) \ot \tilde a \ .
\]
By composing the coaction with the
nontrivial charcter of $A(\IZ_2)$, we find that the generator of $\IZ_2$
acts on $M(2,\IC)$ as the inversion 
$\tilde x \mapsto - \tilde x, ~\tilde y \mapsto - \tilde y$, ie. via an
inner automorphism by
\beq
{\bf U} = \pm \left( \ba{cc}  0 & -1 \\ 1 & 0 \ea \right)~.
\eeq
This action preserves $\IH$ and amounts to the complex conjugation of 
$\alpha$ and of $\beta$ in (\ref{IH}).\\
Next, as far as the algebra $\IC \equiv M(1,\IC )$ is concerned, 
it leads, obviously, to a trivial group $\{e\}$.

We remark also that, since we can embed 
$M(3,\IC )\oplus M(2,\IC ) \oplus \IC$ 
(e.g. in a diagonal way) in
$M(6,\IC )$, 
we have also checked possible quantum symmetries
of $M(6,\IC )$. Repeating our construction, 
there is a quotient Hopf algebra of \slq, $q=e^{2\pi\imath /6}$, 
modulo the relations
\beq\label{F6} 
a^6 = 1 = d^6, ~~ b^3 = 0 = c^3. 
\eeq
It is, however, not interesting 
(even neglecting the problem of invariance of 
the subalgebra $M(3,\IC )\oplus M(2,\IC ) \oplus \IC$)
for the reason that its dimension is 54, 
i.e. just the dimension of $A(F\times \IZ_2) = A(F)\otimes A(\IZ_2)$.

(Incidentally, finding the biggest quantum group coacting  
on a given quantum space seems to be quite an intricate problem).

Obviously, the coaction $\rho_F$ can be extended to the whole $\cal A$ 
in a trivial way by 
\beq 
\label{exte}
\hat{\rho} (m+u+l)= m_{(0)}\otimes m_{(1)}
+ u \otimes 1 + l\otimes 1~,
\eeq
(where we have used Sweedler notation: $\rho_F(m)=m_{(0)}\ot m_{(1)}\in
M(3,\IC)\ot \af$), 
for all $m \in M(3,\IC)$ , $u \in \IH$, $l \in \IC$, so that $\cal A$ becomes 
an \af-comodule algebra. 

A less trivial extension should involve a coaction of another Hopf algebra  
on $\IH$. So far the only candidate we know is $A(\IZ_2)$,
in which case it gives rise to the right coaction of 
$\af \otimes A(\IZ_2)$
\beq \label{bigco} 
\check{\rho} (m+u+l)= m_{(0)}\otimes m_{(1)}\otimes 1
+ u_{(0)}\otimes 1\otimes u_{(1)}  + l \otimes 1\otimes 1~.
\eeq  
With this definition, $\cal A$
becomes an $\af \otimes A(\IZ_2)$-comodule algebra.

\section{Duality}
In the paper \cite{c-r} a parallel approach to quantum finite
symmetries has been discussed in terms of universal enveloping algebras.
The 27-dimensional Hopf algebra ${\cal H}$ defined therein is a quotient Hopf
algebra
of $U_q(sl(2))$ for $q^3=1$. It is generated by $X_+$, $X_-$, $K$, with
the following relations:
\bea 
&& X_+^3=X_-^3=0~, ~~~ K^3=1~,~~~ KX_\pm =q^{\mp 2} X_\pm K~,~~~
 [X_+,X_-] = \frac{K-K^{-1}}{q^{-1}-q}~.
\eea
(Notice that with respect to the convention used by \cite{c-r} we have
changed $q$ with $q^{-1}$, in order to be consistent with our definition
of \slq ).\\
$\cal H$ is a Hopf algebra with coproduct, counit and antipode induced by 
$U_q(sl(2))$: 
\bea
&&\Delta(X_+)= X_+ \ot 1 +K \ot X_+~,~~ \Delta(X_-)= X_- \ot K^{-1} 
+1\ot X_-~,~~\Delta(K) =K\ot K~, \nonumber\\
&& \varepsilon(X_+) =\varepsilon(X_-)=0~,~\varepsilon(K)=1~, \\
&& S(X_+)=-K^{-1}X_+~, ~S(X_-)=-X_-K~,~S(K)=K^{-1}.\nonumber
\eea
In \cite{c-r} it is shown that the underlying vector space of $\cal H$ 
has an intriguing splitting as a direct sum of the semisimple subalgebra
$M(3,\IC) \oplus M(2,\IC) \oplus \IC$,
that is very close to Connes' finite algebra  ${\cal A}$, and the radical
ideal $\cal R$. The radical $\cal R$ is the intersection
of kernels of all 
irreducible representations of \coq~ and it is isomorphic with the algebra of
$3\times 3$ matrices of the form
\[ 
\left(\begin{array}{ccc} 
\alpha_{11} \theta_1\theta_2
& \alpha_{12} \theta_1\theta_2 & \beta_{13} \theta_1 + \gamma_{13}\theta_2
\\
\alpha_{21} \theta_1\theta_2
& \alpha_{22} \theta_1\theta_2 & \beta_{23} \theta_1 + \gamma_{23}\theta_2
\\
\beta_{31} \theta_1 + \gamma_{31}\theta_2
& \beta_{32} \theta_1 + \gamma_{32}\theta_2 & \alpha_{33} \theta_1\theta_2
\end{array}\right) \]
where $\theta_1$, $\theta_2$ are two Grassman variables satisfying the relations 
$\theta_1^2=\theta_2^2=0$ and 
$\theta_1\theta_2 =-\theta_2\theta_1$.\\
It is known \cite{k}, and it is explicit in this presentation, that,
modulo equivalence, there are 
only three irreducible representations of $\cal H$, respectively of
dimension 1, 2 and 3, and that there are no irreducible representations  
of dimension greater than 3.
 From a physical point of view, the basic multiplets of representations of
\coq~ are then,
respectively, a singlet with an arbitrary value of hypercharge, null
isospin and no color, a doublet of 
isospin with zero hypercharge and color, and a triplet of color with zero
isospin and hypercharge. They cannot describe any physical particle, and
using tensor products of basic representations of $\cal H$ via iterating
the coproduct doesn't solve the problem, 
since, as stressed in \cite{c-r}, the subalgebra 
$M(3,\IC) \oplus M(2,\IC)\oplus \IC$ 
is not a subcoalgebra, so that the physical "charges" are not 
additive.
Representations of \coq~ are in general not totally 
reducible, becoming so only when restricted to the semisimple part. In the sequel,
we will give some examples of representations of this kind.

Connes' formulation of the Standard Model uses a 90-dimensional (three
families of leptons and quarks are considered, together with their
antiparticles) representation of ${\cal A}_F$, using the embedding of
$\IH$ in $M(2,\IC)$, so that it is actually obtained by a representation
of the semisimple part of \coq~ (for a good reference, see \cite{m-g-v};
see
also
\cite{l-m-m-s} for problems of such a formulation). Such a representation
can be trivially extended to the whole \coq, by setting to 0 the action
of the radical $\cal R$. It is an open question whether such an extension
is unique.

As regards the relationship between $\cal H$ and \af, 
it is well known that there is a Hopf duality between \slq ~and
$U_q(sl(2))$, in the
sense that there exist a bilinear form $<~,~>$ on $U_q(sl(2)) \times \slq$
such that, for any $u$, $v$ in $U_q(sl(2))$ 
and for any $x$, $y$ in \slq~ one has:
\bea \label{pair}
&& <uv,x> =<u,x_{(1)}><v,x_{(2)}>~,~~
   <u,xy> =<u_{(1)},x><u_{(2)},y>~,\nonumber \\
 &&<1,x>=\varepsilon(x)~,~~<u,1>=\varepsilon(u)~,~~<S(u),x>=<u,S(x)>~.
\eea
Explicitly, this pairing makes use of the fundamental
representation of $U_q(sl(2))$ given by: 
\beq
\rho(X_-) =  \left( \ba{cc} 0 & 1 \\  0 & 0 \ea \right)~,~~
\rho(X_+) =  \left( \ba{cc} 0 & 0 \\  1 & 0 \ea \right)~,~~
\rho(K) =  \left( \ba{cc} q & 0 \\  0 & q^{-1} \ea \right)~.
\eeq
Writing for any $u\in U_q(sl(2))$ \[ \rho(u) =  \left( \ba{cc} A(u) &
B(u) \\ C(u) & D(u) \ea \right)~,\] 
one sets: \[ <u,a>=A(u)~,~ <u,b>= B(u)~,~<u,c>=C(u)~,~ <u,d>= D(u)~,\]
and then extends the definition to arbitrary elements of \slq ~by 
using the properties (\ref{pair}).\\
It turns out that for $q^n=1$ the pairing is degenerate and has a huge kernel.
In particular, for $q^3=1$ the kernel contains both the defining ideals of
the Hopf
algebras \af~ and $\cal H$, so that the pairing descends to the quotients. 

It is convenient to analyse the $27\times 27$ matrix of 
this pairing using as a basis for \af~ the more {symmetric} set 
$\tilde{a}\tilde{b}^r\tilde{c}^s,~\tilde{b}^r\tilde{c}^s,~
\tilde{d}\tilde{b}^r\tilde{c}^s$; $r,s\in\{0,1,2\}$.
Setting ${\mbox deg}(X_-)={\mbox deg}(\tilde{b}) =-1$, ${\mbox
deg}(K)={\mbox deg}(\tilde{a}) ={\mbox deg}(\tilde{d}) =0$, ${\mbox deg}
(X_+) = {\mbox deg}(\tilde{c})=1$, it turns out that monomials with 
different total degree are orthogonal, generating a block diagonal matrix 
with five diagonal blocks.
In table 1 of the Appendix we present those non vanishing blocks.

Now, the determinant of our $27\times 27$ matrix is given by the product 
of the determinants of nine $3\times 3$ subblocks on the diagonal.
It is easy to convince oneself,
by looking at the linear independence of the rows (or the columns) of 
these sub-blocks, that the determinant is different from 0, so that 
the pairing between $\cal H$ and \af~ is not degenerate. 
We are thus in a position to state that 
$\cal H$ and \af~ are dual Hopf algebras.

\section{A representation of \coq~ on $M(3,\IC)$}
Having exhibited the duality between $\cal H$ and \af, 
since $M(3,\IC)$ is a right \af-comodule algebra via the coaction $\rho_F$, 
it becomes a left \coq-module algebra, in the sense that there is a representation
(left action) of \coq~ on $M(3,\IC)$ given by $ h\rhd m
=m_{(0)}<h,m_{(1)}>$, such that 
$h\rhd 1=\varepsilon(h) 1$ and $h\rhd(m m')=
(h_{(1)}\rhd m) (h_{(2)}\rhd m')$.\\
 From these properties, it follows that the generator $K$ acts on $M(3,\IC)$ 
by an automorphism, whereas $X_\pm$ act as {\it twisted} derivations. 
On the basis
of $M(3,\IC)$ given by 
$\tilde{x}^r\tilde{y}^s,~r,s\in\{0,1,2\}$, 
the action of generators $X_{\pm},~K$ is given by:
\bea
&& K\rhd(\tilde{x}^r\tilde{y}^s)= q^{r-s} ~\tilde{x}^r\tilde{y}^s, 
\nonumber \\
&& X_+\rhd(\tilde{x}^r\tilde{y}^s)= \frac{q^r-q^{-r}}{q-q^{-1}}~
\tilde{x}^{r-1}\tilde{y}^{s+1},  \\
&& X_-\rhd(\tilde{x}^r\tilde{y}^s)=\frac{q^s-q^{-s}}{q-q^{-1}}~
\tilde{x}^{r+1}\tilde{y}^{s-1},\nonumber
\eea 
where the summations in exponents are meant modulo 3 and where 
repeated indices are not to be summed on. It is easy to see that there are 
three 3-dimensional invariant subspaces, generated respectively by 
$\{\tilde x^2, \tilde x\tilde y, \tilde y^2\}$, 
$\{\tilde x, \tilde y, \tilde x^2\tilde y^2\}$, 
$\{1,\tilde x^2 \tilde y, \tilde x\tilde y^2\}$, such that on the first 
one \coq~ acts irreducibly, whereas the last two are reducible
indecomposable representation spaces. 

Since $M(3,\IC)$ is simple, the action of $K$ is an inner automorphism,
given in fact as the adjoint action of e.g. 
$\widetilde{K}= \tilde{x}^2\tilde{y}^2$ and 
corresponding to the matrix $U_1$ in eq.(\ref{matrice}).
In addition, the action of $X_{\pm}$ as twisted derivations
can be also expressed as a particular kind of internal operations.
Indeed, $M(3,\IC)$ can be viewed as a $\IZ_3$-graded algebra 
with the grade of the monomials
$m=\tilde{x}^r\tilde{y}^s$ being given by $|m| =r-s {\rm ~mod~} 3$.\\
Then on any element $m$ of grade $|m|$ we have 
\bea
&& X_+\rhd m  = 
\widetilde X_+ ~m - q^{|m|} ~m~ \widetilde X_+, \nonumber \\
&& X_-\rhd m  = 
q^{-|m|} ~\widetilde X_-~ m - m~ \widetilde X_-~,
\eea 
where
\beq
\widetilde X_+=\frac{\tilde{x}^2\tilde{y}}{q^{-1}-q} + C_+ \tilde x^2
\tilde y^2,~~~
\widetilde X_-=\frac{\tilde{x}\tilde{y}^2}{q - q^{-1}}
 + C_- \tilde x^2 \tilde y^2~,
\eeq
with $C_+$, $C_-$ being arbitrary constants.

Note that as elements of $M(3,\IC)$, 
$\widetilde{K}$ and $\widetilde X_\pm$ do not obey 
exactly the same commutations rules of $K$ and $X_\pm$ in \coq. 
For example, to get $\widetilde X_\pm ^3=0$ one can set the constants 
$C_+=\frac{1}{q-q^{-1}}$, $C_-=\frac{1}{q^{-1}-q}$,
but, with this choice one has 
$\widetilde{K}  \widetilde X_\pm \neq q^{\mp 2} \widetilde X_\pm
\widetilde{K}$,
and so on.

We remark that by dualizing (\ref{exte}), we can extend this representation 
of \coq~ on $M(3,\IC)$ to a representation  on $\cal A$, obtaining
\beq 
h\rhd (m+u+l)=
m_{(0)}<h,m_{(1)}>+ (u + l) \varepsilon(h)~,
\eeq
for any $m\in M(3,\IC)$, $u\in \IH$ and $l \in \IC$.\\
Also, in the same way we obtain from (\ref{bigco}) a representation of 
$ {\cal H} \otimes \IC(\IZ_2)$ on  $\cal A$, where $\IC(\IZ_2)$ is the
group
algebra of $\IZ_2$, which is in a natural duality with $A(\IZ_2)$ 
\cite{m-s}. 
Explicitly, the action of a simple tensor $h\otimes z $
in ${\cal H}\otimes \IC(\IZ_2)$ on an element $m +u+l$ turns out to be
\beq 
h\otimes z \rhd (m+u+l)=
m_{(0)}<h,m_{(1)}>\varepsilon(z) + u_{(0)}<z,u_{(1)}>\varepsilon(h) + l
\varepsilon(h)\varepsilon(z)~.
\eeq

\section{Further properties of \coq~ and \af}
Using again duality we can compute two different commuting representations 
of \coq~ on \af~. One of them is given by $<R(h)(\phi),h'>=<\phi,h'h>$, 
or in Sweedler notation 
\beq
 h\rhd\phi=\phi_{(1)}<h,\phi_{(2)}>. 
\eeq 
Such a representation, which corresponds by duality \cite{m-s} to 
the comultiplication in \af, makes \af~ a left-\coq~ module algebra. 
In table 2 of the Appendix we show the values of the action of the generators 
of \coq~ via such representation on the basis of \af.\\
The other representation is given by
$<L(h)(\phi),h'>=<\phi,S(h)h'>$, 
or in Sweedler notation
\beq
 h\rhd\phi=<S(h),\phi_{(0)}>\phi_{(1)}. 
\eeq
The representation $L$ is such that $h\rhd (\phi\psi)=(h_{(2)}\rhd\phi)
(h_{(1)}\rhd\psi)$, $h\rhd 1 =\varepsilon(h)$, and corresponds to the
right coaction
of \af~ on itself given by $\Delta_R =(id \otimes S)\circ \tau\circ 
\Delta$, where 
$\tau$ is the flip operator. In table 3 of the Appendix we present 
explicitly the action of generators.

Recall now that a left (resp. right) integral {\it on} a Hopf algebra $H$
over a field $k$ 
is a linear functional $h:H \ra k$ satisfying $
(Id \ot h) \circ \Delta = 1\sb H \cdot h ~~~
(\mbox { resp. } (h\ot Id)\circ \Delta = 1\sb H \cdot h )$, 
whereas an element $\lambda \in H$ is called a left (resp. right) integral 
{\it in} $H$ if it verifies $\alpha\lambda =\varepsilon(\alpha)\lambda$,
or, respectively, 
$\lambda\alpha =\varepsilon(\alpha)\lambda$, for any $\alpha \in H$.
For finite dimensional Hopf algebras, integrals {\it in} $H$ are nothing
but integrals {\it on} the dual $H^*$ and both the spaces of 
left and right integrals are one dimensional \cite{s}.
A Hopf algebra $H$ is called
{\it unimodular} if there are left and right integrals on $H$ which coincide.
An integral on unimodular $H$ is called a {\it Haar measure} 
iff it is {\it normalized}, i.e.  $h (1)= 1$.

In our case we have that
the Hopf algebra \af ~is unimodular with the left and right integrals 
being given by $h= C(\tilde{b}^2\tilde{c}^2)^*$, i.e. 
$h (\tilde{b}^2\tilde{c}^2) = ~C\in k~$,
$h (\tilde{a}^p \tilde{b}^r\tilde{c}^s) = 0$ if $(p, r, s) \neq (0, 2, 2)$
\cite{dhs}.
In terms of the basis 
of ${\cal H} =\af^*$, $h= C X_-^2X_+^2(1+K+K^2)$.
Being $h$ not normalizable, it follows that there is no Haar measure on
the Hopf algebra \af ~and, consequently, $F$ is not a 
compact matrix quantum group in the sense of \cite{w-s}.\\ 
As regards integrals {\it in} \af, it is easy to see that the
element 
$\lambda_L =(1+\tilde{a}+\tilde{a}^2)
\tilde{b}^2\tilde{c}^2$ is a left integral and that the element  
$\lambda_R =
\tilde{b}^2\tilde{c}^2 (1+\tilde{a}+\tilde{a}^2)$ is a right one.
Thought as integrals {\it on} $ \cal H$, $\lambda_L=(X_-^2X_+^2K)^*$
 and $\lambda_R=(X_-^2X_+^2K^2)^*$.  
Thus in this case left and right integral are not proportional. It is
evident, now, that, as stated in Proposition 7 in \cite{ls}, there exist 
(left and right) integrals in and on \af~ and \coq~ such that $<h,\lambda>=1$. 

In addition, by Theorem 5.18 in \cite{s},
the property $\varepsilon(\lambda_L)= \varepsilon(h)=0$ assures us that
both \af ~and {\cal H} are neither semisimple nor cosemisimple. 

 \section*{Acknowledgements} 
It is a pleasure to thank Piotr Hajac for helpful advices.   

 \section*{Appendix: Explicit Duality and Representations}  
\def\E{X_-}
\def\F{X_+}
\def\K{K}
\def\a{\tilde a}
\def\b{\tilde b}
\def\C{\tilde c}
\def\d{\tilde d}
\def\rrr{\vrule height 3ex depth 1.4ex width 0cm}  
\def\xxx#1{\hbox to 3ex{\hss$\scriptstyle#1$\hss}}
\def\yyy#1{\rrr{\scriptstyle#1}}

\begin{table}[ht]
\begin{tabular}{cc}
$\begin{array}{r||ccc|ccc|ccc}
\left<\cdot|\cdot\right>&
\xxx{\b^2\C^2}&\xxx{\a\b^2\C^2}&\xxx{\d\b^2\C^2}&\xxx{\b\C}&
\xxx{\a\b\C}&\xxx{\d\b\C}&\xxx{1}&\xxx{\a}&\xxx{\d}\\
\hline\hline
\yyy{\E^2\F^2 }&1&q^2&q&0&q&0&0&0&0\\
\yyy{\E^2\F^2\K }&1&1&1&0&q^2&0&0&0&0\\
\yyy{\E^2\F^2\K^2 }&1&q&q^2&0&1&0&0&0&0\\
\hline
\yyy{\E\F }&0&0&0&1&q&q^2&0&1&0\\
\yyy{\E\F\K }&0&0&0&1&q^2&q&0&q&0\\
\yyy{\E\F\K^2 }&0&0&0&1&1&1&0&q^2&0\\
\hline
\yyy{ 1}&0&0&0&0&0&0&1&1&1\\
\yyy{ \K}&0&0&0&0&0&0&1&q&q^2\\
\yyy{ \K^2}&0&0&0&0&0&0&1&q^2&q
\end{array}$&
~~~\begin{tabular}{c}
$\begin{array}{r||ccc}
\left<\cdot|\cdot\right>&
\xxx{\C^2}&\xxx{\a\C^2}&\xxx{\d\C^2}\\
\hline\hline
\yyy{\F^2 }	&-1&-q^2&-q\\
\yyy{\F^2\K }	&-q^2&-q^2&-q^2\\
\yyy{\F^2\K^2}	&-q&-q^2&-1\\
\end{array}$\\~\\
$\begin{array}{r||ccc}
\left<\cdot|\cdot\right>&
\xxx{\b^2}&\xxx{\a\b^2}&\xxx{\d\b^2}\\
\hline\hline
\yyy{\E^2 }	&-1&-1&-1\\
\yyy{\E^2\K }	&-q&-q^2&-1\\
\yyy{\E^2\K^2}	&-q^2&-q&-1\\
\end{array}$
\end{tabular}
\end{tabular}

\bigskip
\begin{tabular}{cc}
$\begin{array}{r||ccc|ccc}
\left<\cdot|\cdot\right>&
\xxx{\b^2\C}&\xxx{\a\b^2\C}&\xxx{\d\b^2\C}&\xxx{\b}&
\xxx{\a\b}&\xxx{\d\b}\\
\hline\hline
\yyy{\E^2\F }&-1&-q&-q^2&0&-1&0\\
\yyy{\E^2\F\K }&-q^2&-q&-1&0&-1&0\\
\yyy{\E^2\F\K^2 }&-q&-q&-q&0&-1&0\\
\hline
\yyy{\E }&0&0&0&1&1&1\\
\yyy{\E\K }&0&0&0&q^2&1&q\\
\yyy{\E\K^2 }&0&0&0&q&1&q^2
\end{array}$&
~~$\begin{array}{r||ccc|ccc}
\left<\cdot|\cdot\right>&
\xxx{\b\C^2}&\xxx{\a\b\C^2}&\xxx{\d\b\C^2}&\xxx{\C}&
\xxx{\a\C}&\xxx{\d\C}\\
\hline\hline
\yyy{\E\F^2 }	&-1&-q^2&-q&0&-q&0\\
\yyy{\E\F^2\K }	&-q&-q&-q&0&-1&0\\
\yyy{\E\F^2\K^2}&-q^2&-1&-q&0&-q^2&0\\
\hline
\yyy{\F }	&0&0&0&1&q&q^2\\
\yyy{\F\K }	&0&0&0&q&1&q^2\\
\yyy{\F\K^2 }	&0&0&0&q^2&q^2&q^2
\end{array}$
\end{tabular}
\caption{Diagonal blocks in the pairing of $\cal H$ and \af}
\end{table}

\begin{table}
\hskip3cm
$\begin{array}{r||ccc}
\hbox{\scriptsize R(h)}&
\xxx{\E}&\xxx{\F}&\xxx{\K}\\
\hline\hline
1&0&0&1\\
\a&0&\b&q\a\\
\d&\C&0&q^2\d\\
\b&\a&0&q^2\b\\
\a\b&\d-q\d\b\C+q^2\d\b^2\C^2&\b^2&\a\b\\
\d\b&1-\b\C&0&q\d\b\\
\b^2&-\a\b&0&q\b^2\\
\a\b^2&-\d\b+q\d\b^2\C&0&q^2\a\b^2\\
\d\b^2&-\b&0&\d\b^2\\
\C&0&\d&q\C\\
\a\C&0&q-q\b\C&q^2\a\C\\
\d\C&q^2\C^2&q^2\a-q\a\b\C+\a\b^2\C^2&\d\C\\
\b\C&q^2\a\C&\d\b&\b\C\\
\a\b\C&q^2\d\C-\d\b\C^2&q\b-q\b^2\C&q\a\b\C\\
\d\b\C&q^2\C+\b\C^2+q\b\C^2&q^2\a\b-q\a\b^2\C&q^2\d\b\C\\
\b^2\C&-q^2\a\b\C&\d\b^2&q^2\b^2\C\\
\a\b^2\C&-q^2\d\b\C+\d\b^2\C^2&q\b^2&\a\b^2\C\\
\d\b^2\C&-q^2\b\C&q^2\a\b^2&q\d\b^2\C\\
\C^2&0&-q\d\C&q^2\C^2\\
\a\C^2&0&-q^2\C&\a\C^2\\
\d\C^2&0&-\a\C+q^2\a\b\C^2&q\d\C^2\\
\b\C^2&q\a\C^2&-q\d\b\C&q\b\C^2\\
\a\b\C^2&q\d\C^2&-q^2\b\C&q^2\a\b\C^2\\
\d\b\C^2&q\C^2&-\a\b\C+q^2\a\b^2\C^2&\d\b\C^2\\
\b^2\C^2&-q\a\b\C^2&-q\d\b^2\C&\b^2\C^2\\
\a\b^2\C^2&-q\d\b\C^2&-q^2\b^2\C&q\a\b^2\C^2\\
\d\b^2\C^2&-q\b\C^2&-\a\b^2\C&q^2\d\b^2\C^2
\end{array}$
\caption{Action of the generators of \coq~ via the representation $R$}
\end{table}

\begin{table}
\hskip3cm
$\begin{array}{r||ccc}
\hbox{\scriptsize L(h)}&
\xxx{\E}&\xxx{\F}&\xxx{\K}\\
\hline\hline
1&0&0&1\\
\a&-q^2\C&0&q^2\a\\
\d&0&-q\b&q\d\\
\b&-q^2\d&0&q^2\b\\
\a\b&-1+\b\C&q&\a\b\\
\d\b&-q\a+\a\b\C-q^2\a\b^2\C^2&-\b^2&\d\b\\
\b^2&\d\b&0&q\b^2\\
\a\b^2&q\b&0&\a\b^2\\
\d\b^2&q^2\a\b-q\a\b^2\C&0&q^2\d\b^2\\
\C&0&-q\a&q\C\\
\a\C&-q^2\C^2&-q\d+q^2\d\b\C-\d\b^2\C^2&\a\C\\
\d\C&0&-q+q\b\C&q^2\d\C\\
\b\C&-q^2\d\C&-\a\b&\b\C\\
\a\b\C&-\C+\b\C^2&-\d\b+q\d\b^2\C&q^2\a\b\C\\
\d\b\C&-q\a\C+\a\b\C^2&-\b+\b^2\C&q\d\b\C\\
\b^2\C&\d\b\C&-q^2\a\b^2&q^2\b^2\C\\
\a\b^2\C&q\b\C&-q^2\d\b^2&q\a\b^2\C\\
\d\b^2\C&q^2\a\b\C-q\a\b^2\C^2&-q^2\b^2&\d\b^2\C\\
\C^2&0&q\a\C&q^2\C^2\\
\a\C^2&0&q\d\C-q^2\d\b\C^2&q\a\C^2\\
\d\C^2&0&q\C&\d\C^2\\
\b\C^2&-q^2\d\C^2&\a\b\C&q\b\C^2\\
\a\b\C^2&-\C^2&\d\b\C-q\d\b^2\C^2&\a\b\C^2\\
\d\b\C^2&-q\a\C^2&\b\C&q^2\d\b\C^2\\
\b^2\C^2&\d\b\C^2&q^2\a\b^2\C&\b^2\C^2\\
\a\b^2\C^2&q\b\C^2&q^2\d\b^2\C&q^2\a\b^2\C^2\\
\d\b^2\C^2&q^2\a\b\C^2&q^2\b^2\C&q\d\b^2\C^2
\end{array}$
\caption{Action of the generators of \coq~ via the representation $L$}
\end{table}

\end{document}